\documentclass[11pt]{article}\topmargin=-0.5cm\hfuzz=10pt \oddsidemargin=0.1cm \textheight=215mm\textwidth=16.3cm 
\usepackage[pagewise]{lineno}
\newcommand{\comm}[1]{}

\usepackage{amssymb}
\usepackage{mathrsfs}
\usepackage{eufrak}
\usepackage{amsmath}\usepackage{graphicx} \usepackage[numbers]{natbib}
\usepackage{epsfig}
\def\citet{\cite}

\def\xxxonly{\comm}%
\def\xxxonly{}
\def\noxxx{\comm}
 \usepackage{color}

\newcommand{\rd}[1]{\color{red}#1\color{black}} 
\newcommand{\bl}[1]{\color{blue}#1\color{black}}

\usepackage{times}\hfuzz=10pt 

   \usepackage{graphicx}
\usepackage{epstopdf}

\newtheorem{theorem}{Theorem}
\newtheorem{lemma}{Lemma}
\newtheorem{proposition}{Proposition}
\newtheorem{corollary}{Corollary}
\newtheorem{condition}{Condition}
\newtheorem{definition}{Definition}
\newtheorem{remark}{Remark}
\newtheorem{example}{Example}

\def\e{\varepsilon}

\def\defi{\stackrel{{\scriptscriptstyle \Delta}}{=}}

\def\d{\delta}
\def\o{\omega}
\def\O{\Omega}

\def\F{{\cal F}}
\def\w{\widehat}
\def\Ind{{\mathbb{I}}}

\def\mes{{\rm mes\,\!}}

\def\esssup{\mathop{\rm ess\,\! sup}}

\def\Re{{\rm Re\,\!}}

\def\R{{\bf R}}

\def\F{{\cal Z}}

\def\L{L}
\def\F{{\Feta}}

\def\C{{\bf C}}

\def\ww{\widetilde}

\def\oo{\bar}

\def\G{\Gamma}
\def\GG{{\cal G}}
\def\U{{\cal U}}

\def\M{{\cal M}}

\def\L{{\cal L}}
\def\I{{\,\! \cal I}}

\def\F{{\cal F}}

\newcommand{\be}{\begin{equation}}
\newcommand{\ee}{\end{equation}}
\newcommand{\bd}{\begin{displaymath}}
\newcommand{\ed}{\end{displaymath}}
\newcommand{\ba}{\begin{array}{ll}}
\newcommand{\ea}{\end{array}}
\newcommand{\baa}{\begin{eqnarray}}
\newcommand{\eaa}{\end{eqnarray}}
\newcommand{\baaa}{\begin{eqnarray*}}
\newcommand{\eaaa}{\end{eqnarray*}}

\def\ZZ{{\bf Z}}

\def\oo{\bar}

\def\CC{{\cal C}}

\def\ew{\left(e^{i\o}\right)}
\def\ew{\left(i\o\right)}

\def\TT{{\cal T}}\def\TT{{\cal T}}

\def\HHH{{\cal H}}
\def\II{{\cal I}}
\def\ee{\epsilon}
\def\TTT{{\mathscr{T}}}
\def\TTT{{\mathfrak{T}}}
\def\M{{\cal M}_{\G,I}}
\def\HH{{\rm H}}
\title{Extrapolation and sampling  for processes on  spatial graphs}
\author{
Nikolai Dokuchaev}
 
\begin{document}
\def\rd{\comm}\def\bl{}
 \vspace{-1.5cm}      \maketitle
\def\brea{}
\def\breakk{}
\def\break{}
\noxxx{\let\thefootnote\relax\footnote{
The author is with Zhejiang University/University of Illinois at Urbana-Champaign Institute,  
Zhejiang University, Haining, China. Email: Dokuchaev@intl.zju.edu.cn. }}\let\thefootnote\relax\footnote{Submitted to arXiv: August 20, 2019. Revised:  May 11, 2022.}
{\let\thefootnote\relax\footnote{
The paper is accepted to "Sampling Theory, Signal Processing, and Data Analysis".}}
\begin{abstract}
The paper studies processes defined on time domains structured as oriented spatial  graphs (or metric graphs, or oriented branched 1-manifolds). This setting can be used, for example, for forecasting models involving
branching scenarios.
For these  processes, a notion of the spectrum degeneracy  that takes into account the topology of the graph is introduced.
The paper suggests sufficient conditions of uniqueness of extrapolation and recovery from the observations on a single branch.
This also implies an analog of sampling theorem for branching processes, i.e.,  criterions of their recovery from
a set of equidistant samples, as well as  from a set of equidistant samples from  a single branch.

\end{abstract}
\section{Introduction}
Models involving processes on spatial graphs and branched manifolds have applications to the description of a number of processes in quantum mechanics and biology. Currently, there are many results for differential equations and stochastic differential equations for the state space represented by  metric graphs; see, e.g., \cite{CP,FM,HH,H1M,H2M,KM,PP,P}, and the literature therein.

This setting can be used, for example, for forecasting models involving
branching scenarios.

 In signal processing on graphs, the main efforts
are directed  toward   the sampling on the vertices in the discrete setting; see, e.g.,
\citet{Anis,Anis2,Chen,Jung1,Jung2,NgD,SM,SN} and the references therein.  The present  paper extends   
basic results for  the continuous time
signal processing  on the case of the time domains structured as oriented
metric graphs with non-trivial topological structure.

More precisely, the paper studies  the  problem of spectral characterization of uniqueness of recovery of a process
from its trace on a branch, in the signal processing setting
based on frequency analysis and sampling.
The existing theory of extrapolation and forecasting covers  processes that do not involve branching.

The framework developed in this paper  could provide new possibilities for sampling and extrapolation for models involving
branching scenarios. Let us provide a basic example of this model.

\begin{itemize}
\item
Let an observer tracks a process  $x(t)$ up to time $t<0$, and, at time $t=0$, the process split  in two processes $x_1(t)$ and $x_2(t)$ for $t\ge 0$; suppose that  $x_1(t)=x_2(t)=x(t)$ for $t<0$.  For example, this is a situation where a locator tracks a fighter jet that ejects a false target.
The classical sampling theorem does not allow to  represent  these processes via discrete equidistant samples,   since it would require that all processes are band-limited, which is impossible
since the band-limited extension of $x(t)$ from the domain $\{t<0\}$  is unique.
 \end{itemize}
 \noindent
As a solution,  we suggest  to consider a set of processes $\{x_k(t)\}$ that all coincide on $(-\infty,0)$
 and that has mutually disjoint spectrum gaps, and that
 the path $x|_{t<0}=x_k|_{t<0}$ allows a number of different unique extrapolations corresponding to different hypothesis
 about the locations of the spectrum gaps for $x_k$. This allows to extend the sampling theorem on this branched process.  The paper applies this approach for  more general setting.

Let us list some basic extrapolation results for the classical setting  without branching.
It is known that there are some   opportunities for prediction and interpolation
of continuous time processes  with  certain degeneracy of
their spectrum.   The  classical Sampling Theorem  states that
 a band-limited continuous time function can be uniquely recovered without error  from  a  sampling sequence   taken with sufficient frequency.  Continuous time functions with periodic gaps for the Fourier transform
can be recovered from sparse samples;
see \cite{La2,OU08,OU}. Continuous time band-limited functions are analytic and can be recovered from the values on an arbitrarily small time interval.
In particular, band-limited functions can be predicted from their past values. Continuous time functions with the Fourier transform vanishing on an arbitrarily small interval $(-\O,\O)$ for some $\O>0$  are also uniquely defined by their past values \citet{D08}.

Spectrum degeneracy is a  quite special feature that is difficult to ensure for a process. However, in many cases, the extrapolation methods being developed  for processes with spectrum degeneracy feature some
robustness  with respect to the noise contamination. In some cases, it is possible to
apply these methods for projections of underlying processes on the space of processes with spectrum degeneracy; see, e.g., \cite{D18}.

It appears that many applications require to extend the existing theory of sampling and extrapolation on the processes
defined on a  time domains with non-trivial structures.
These structures may appear for hybrid dynamic systems,   with regime switches; see, e,g. \cite{SS}.
 There are also models for partial differential  systems  with the state space represented by  branched manifolds see, e.g., \cite{CP,HH,PP,P}, and the literature therein.

The paper suggests a simple but effective  approach allowing to use the standard Fourier transform for the traces  on  sole  branches that are deemed to be extended  onto the entire real axis.
The topology of the system is taken into account via a  restriction that these branch processes (or their transformations) coincide on preselected parts of the real axis; the selection of these parts
and transformations  defines  the topology of the branched 1-manifold  representing the time domain.
This approach allows  a relatively simple and convenient
representation of processes defined on time represented as a 1-manifold, as well as more general processes
described via  restrictions such  as
$x_1(t)= x_2(t)$ for $t<a$, or $x_3(t)=\int_\R h(t-s)x_4(s)ds$ for $t\in (b,c)$, with  arbitrarily chosen  preselected  $a,b,c\in\R$ and functions $h:\R\to \R$.

The paper suggests sufficient conditions of uniqueness of extrapolation and recovery from the observations on a single branch
and from a set of equidistant samples from  a single branch.
It appears that the processes spectrum degeneracy of the suggested kind are everywhere dense in a wide class of the underlying processes, given some restrictions
 on the topology of the underlying 1-manifold   (Lemma\ref{ThM}).  Some applications to extrapolation and sampling are considered.
In particular, it is shown  that a
 process defined on  a time domain structured as a  tree allows an arbitrarily close approximation by a function
that is uniquely defined by its equidistant sample
 taken on a semi-infinite half of a root branch (Corollary \ref{corrS2}).

\xxxonly{A related work \cite{D17firstversion} considers a similar but significantly  simpler model for the time domain.}

The paper is organized in the following manner.
 In Section \ref{SecD},  we provide some definitions and an adaptation of known results on uniqueness of
 extrapolation in the standard time domain.
  In Section \ref{SecM}, we provide the main results on conditions  of uniqueness of extrapolation of a branched
  process from a sample taken from a single branch and conditions of possibility of approximation
 of a general type branched process by processes allowing the unique extrapolation
 (Lemma \ref{ThU} and Theorem \ref{ThM}).  Section \ref{SecP} contains the proofs.  Section \ref{SecC} offers some discussion and concluding remarks.

\section{Definitions and some background facts}\label{SecD}
\rd{We denote by $L_p(\R)$ the standard space of complex valued functions $L_p(\R,\C)$, $p\in[1,+\infty]$. }

For complex valued functions $x\in L_1(\R)$ or $x\in L_2(\R)$, we
denote by $\F x$ the function defined on $i\R$, where $i=\sqrt{-1}$, as the Fourier
transform  $$(\F x)(i\o)= \int_{-\infty}^{\infty}e^{-i\o
t}x(t)dt,\quad \o\in\R.$$ If $x\in L_2(\R)$, then $X(i\cdot)$ is defined
as an element of $L_2(\R)$ (meaning that  $X\in L_2(i\R)$). If $X(i\cdot)\in L_1(R)$ then
$x=\F^{-1}X\in C(\R)$ (i.e. it is a bounded and continuous function on $\R$).

A process $x\in L_2(\R)$ is said to be band-limited if its Fourier transform has a bounded support on $i\R$.
 These processes can be recovered from their equidistant samples; the required frequency of sampling depends
  on the size of this support.

Let $m>0$ be a fixed integer.

Let $\I_f$ be the set of all Borel subsets of $\R$ of positive finite measure.
Let $\I_\infty$ be the set of  all Borel subsets $B\subset \R$ such that there exists $a\in\R$ such that
either $(a,+\infty)\subset B$ or $(-\infty,a)\subset B$. Let $\I=\I_f\cup \I_\infty$.

 Let $\G$ be  a set $\{(d,k)\}$ of ordered pairs  such that
 $d,k\in\{1,...,m\}$, $d\neq k$ and that if $(d,k)\in \G$ then
 $(k,d)\notin \G$.
 The set  $\G$ is  non-ordered.

We assume that, for each $\G$,  there is a mapping $I:\G\to \I$. We denote $I_{k,d}=I(k,d)$.
\rd{By the definitions, it follows that $I_{d,k}=I_{k,d}$.}

Let $\HHH$ be the set of all  sets  $h=\{h_{d,k}\}_{(d,k)\in\G}$,
where $h_{d,k}:L_2(\R)\to L_2(\R)$ are continuous mappings.

Let $\TTT$ be the set of all triplets $\TT=(\G,I,h)$, where $I\in\I$ and  $h\in\HHH$.

\begin{definition} \label{def1} \begin{enumerate}
\item
For a given $\TT\in\TTT$ , let $\L_{2,\TT}$ be the set of all ordered
 sets $\{ x_d\}_{d=1}^{m}\in [L_2(\R)]^m$ such that
  $x_k|_{I_{d,k}}=h_{d,k}(x_d)|_{I_{d,k}}$
  for all $(d,k)\in\G$ up to equivalency, i.e.  $x_k(t)=(h_{d,k}x_d)(t)$ almost everywhere on $I_{d,k}$.
\item For a given $\TT\in\TTT$ , let $\CC_{\TT}$ be the set of all  $\{ x_d\}_{d=1}^{m}\in \L_{2,\TT}$ such that $x_d\in C(\R)$
and  $X_d(i\cdot)\in L_1(\R)$ for all $d$, where
 $X_d=\F x_d$.
 \end{enumerate}
In all these cases, we say that $\{ x_d\}_{d=1}^{m}$ from Definition \ref{def1}  is  a $\TT$-branched process.
\end{definition}

Let us discuss the connection of Definition \ref{def1} with the setting for processes defined on oriented branched 1-manifolds.

\par
Consider first the case
where $h_{dk}$ is the identity operator for all $(d,k)\in\G$, i.e.,  $h_{dk}(x)\equiv x$.
In this case, each pair $(\G,I)\in \GG\times \II$ can be  associated with  a branched 1-manifold $\M$
formed as the union of $m$ infinite lines  representing usual 1D time, with coordinates $t_1,...,t_m$, respectively,
such that, for all $(d,k)\in\G$, line $d$ and line $k$ are "glued" together at $I_{d,k}$, i.e., $t_d=t_k$ if $t_d\in I_{d,k}$ and  $t_k\in I_{d,k}$.
Hence, in this case, a $\TT$-branched process  $\{ x_d\}_{d=1}^{m}$ can be identified with a process $x:\M\to \C$.
Therefore, a special case of  Definition \ref{def1} with trivial $h$  provides a  convenient
representation of processes defined on branched 1-manifold $\M$.

For the case of  non-trivial  choices of $h$,   Definition \ref{def1} leads to additional opportunities
for modeling processes that are mutually connected by non-trivial ways, such as  $x_k(t)=x_d(a t+b)+c$ or
  $x_k(t)=\int_\R h(t-s)x_d(s)ds$ for $t\in I_{d,k}$ for any $a,b,c\in\R$ and $h\in L_2(\R)$.
In this more general case, $\TT$-branched processes cannot be described as just functions $x:\M\to \C$.

\begin{figure}[ht]
\centerline{\psfig{figure=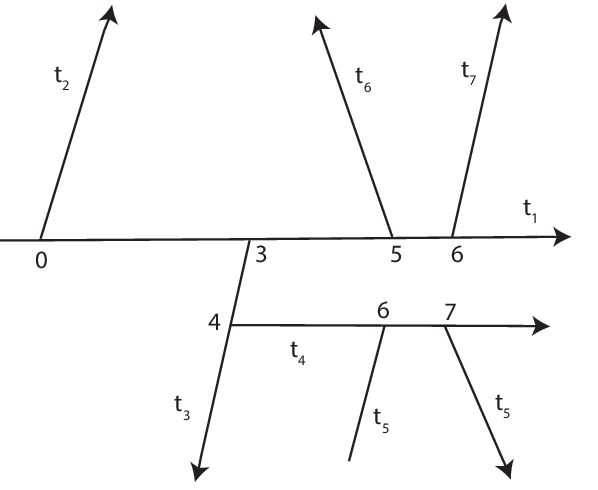,width=9cm,height=5.5cm}}
\caption[]{ The structure of the manifold  $\M$ for Example  \ref{exA}.}
\label{fig1}
\end{figure}

\begin{example}\label{ex1} {\rm
Let an observer  tracks a fighter jet at times $t<0$, and the jet  ejects a false target at time $t=0$.
The classical sampling theorem does not  applicable to this case of branching paths.
Consider a function on a 1-manifold
that can be associated with a $\TT$-branched process. Let $\M$ be Y-shaped manifold with one infinite branch and  one semi-infinite branch. Let $(t_1,t_2)\in\R\times (0,+\infty)$ be
coordinates  for the first and second branches, respectively. Assume that the branching  point is located at $(t_1,t_2)=(0,0)$.
  Let $y:\M\to\C$ be a function.
The process $y$  of 1-manifold can be represented via a $\TT$-branched process
with $m=2$ and $\TT=((1,2),I,h)$,
where $I_{1,2}=(-\infty,0)$, $h_{1,2}:L_2(\R)\to L_2(\R)$ is the identity operator, i.e., $h_{1,2}(x)\equiv x$.
The corresponding $\TT$-branched process $\{x_d\}_{d=1,2}$ is such that $x_1(t_1)=y(t_1)$ for all $t_1$,
 $x_1(t_1)=x_2(t_1)$ for all $t_1<0$, and  $x_2(t_2)=y(t_2)$ for all $t_2>0$. }
\end{example}
\begin{example}\label{exA}
Consider $\TT$-branched process
$\G=\{(1,2),(1,3),(1,6), (1,7),(3,4),(4,5)\}$, and $I:\G\to \I$ such that
\baaa
&&I_{1,2}=(-\infty,0), \quad I_{1,3}=(3,+\infty),\quad I_{1,6}=(5,+\infty),\quad
\breakk I_{1,7}=(-\infty,6),
\quad
 I_{3,4}=(-\infty,4), \quad I_{4,5}=(6,7).
 \eaaa Assume that  $h_{dk}(x)=x$ for all $(d,k)\in\G$ except $(1,3)$,
and where  $h_{13}(x)(t)=x(6-t)$,  this would correspond to
 restrictions $x_1(t)=x_2(t)$ for $t<0$,  $x_1(t)=x_3(6-t)$ for $t>3$,
 $x_4(t)=x_3(t)$ for $t>4$, $x_5(t)=x_4(t)$ for $t\in (6,7)$.
 This branched 1-manifold $\M$  is represented by Figure \ref{fig1}.
 \end{example}

  \begin{definition}\label{defDEG}
  \begin{enumerate}
  \item  We denote by $\GG$ the set of all ordered sets $G=(G_1,..,G_m)$, where $G_d\in\I$, $d=1,...,m$.  We denote by $\oo\GG$ the set of all ordered sets $G=(G_1,...,G_m)$, where
  either $G_d=\emptyset$ or $G_d\in\I$, $d=1,...,m$.
   \item
 For $G\in \oo\GG$, we denote by
$\L^G_{2,\TT}$ the set  of  all  $\TT$-branched processes $\{x_d\}_{d=1}^m$ from $\L_{2,\TT}$
such that
$X_d\ew=0$ for $\o\in G_d$, where $X_d=\F x_d$.
 \end{enumerate}
 \end{definition}
One may refer  $G_d$ as the spectrum gaps of $x_d$.
\begin{proposition}\label{lemmaU}
For $\oo I\in\I$ and $\oo G\in \I$, let  $\U_{\oo I,\oo G}$ be the set of all  $x\in  L_2(\R)$
such that  \rd{ $x(t) =0$ for $t\in \oo I$ and } $X\ew =0$ for $\o\in \oo G$, where  $X=\F x$.
Let  $\mes(\oo I\cup \oo G)=+\infty$.
Then any $x\in U_{\oo I,\oo G}$ is uniquely defined by its path $x|_{\oo I}$.
\end{proposition}
\begin{corollary}\label{lemmaU2}
Let $\TT=(\G,I)\in \TTT$ and  $\{x_d\}_{d=1}^m\in \L^G_{2,\TT}$, where $G=(G_1,...,G_m)\in\GG$.
Let $\{d,k\}\in\G$ and $G_d\in \I$, $G_k\in \I$.
\begin{enumerate}
\item
If \bl{$\mes(I_{d,k}\cup G_{k})=+\infty$,} then $x_k$ is uniquely defined by the path $h_{d,k}(x_d)|_{I_{d,k}}$.
\item
If $I_{d,k}\in \I_\infty$ and $\mes (G_d\cap G_k)>0$, then $x_k\equiv h_{d,k}(x_d)$.
\item
If  $\mes(G_d\cap G_k)=+\infty$, then  $x_k\equiv h_{d,k}(x_d)$.
\end{enumerate}
\end{corollary}
\def\eqq{\rightharpoonup}
\begin{definition}\label{defBranch} Let $\TT=(\G,I,h)\in\TTT$.
Let $d_0,d\in \{1,...,m\}$, $d_0\neq d$. We say that  $d_0\eqq d$    if there exists a sequence $\{d_1,...,d_j\}\subset \{1,...,m\}$
 such that \baa
 &&(d_0,d_1),\  (d_1,d_2),\ \cdots,\   (d_{j},d)\in\G,\nonumber \\
 &&
 I_{d_0,d_1},\ \ I_{d_1,d_2},\ \ \cdots,\ \  I_{d_{j},d}\in I.
 \label{sim}\eaa
  \end{definition}
It can be noted that  if $h_{d,k}(x)=x$ for all $(d,k)\in \G$ then the relation $\eqq$  is symmetric.
 \section{Main results}\label{SecM}
Let us state first some conditions allowing to recover the entire $\TT$-branched process form a single branch.

\begin{lemma}\label{ThU}
Let $\TT$ and  $G=(G_1,...,G_m)$ be given  such that
$G_d\in\I$ for $d\ge 2$.
Assume  $1\eqq d$ for any $d\in \{2,...,m\}$ such that (\ref{sim}) holds
 and such that
\baa
\mes ( I_{d_{k-1}} \cup G_{d_k} ) =+\infty, k=1,...,j.
\label{IG}
\eaa
(We assume that $1=d_0$ and $d=d_{j+1}$  in (\ref{sim})).
\index{\baa
&&\hbox{if}\quad  I_{d_{k-1},d_k}\in \I_f\quad\hbox{then}\quad G_{d_k}\in \I_\infty,
\label{IG1}\\
&&\hbox{if}\quad G_{d_k}\in \I_f\quad\hbox{then}\quad I_{d_{k-1},d_k} \in \I_\infty.
\label{IG2}
\eaa}
Assume that  $\{x_d\}_{d=1}^m\in \L^G_{2,\TT}$. Then
\begin{enumerate}
\item
$\{x_d\}_{d=1}^{m}$
is uniquely defined by  $x_1$;
\item
If $G_1\in \I$ then $\{x_d\}_{d=1}^{m}$
is uniquely defined by the path $x_1|_{\oo I}$, for any $\oo I\in \I_\infty$;
\item
If $G_1\in \I_\infty$ then
$\{x_d\}_{d=1}^{m}$
is uniquely defined by the path $x_1|_{\oo I}$, for any $\oo I\in \I$.
\end{enumerate}
\end{lemma}

We say that a $\TT$-branched process
$\{ x_d\}_{d=1}^{m}$ such as described in Lemma \ref{ThU} features  branched spectrum  degeneracy with the parameter $(\TT,G)$.

It can be clarified that, in Lemma \ref{ThU}, the components  $x_d$ are defined uniquely in $L_2(\R)$.
However, for  a  $\TT$-branched process  from $ \C_{\TT}$,  the components  $x_d$ are  defined uniquely  in $C(\R)$.
\begin{remark}
It can be noted that the degeneracy required in Lemma \ref{ThU} can be arbitrarily small, i.e.,
 $\mes(G_{d_k})$ can be arbitrarily  small under assumption (\ref{IG}) given that $I_{d,k}\in\I_\infty$.
\end{remark}
\begin{remark} Lemma \ref{ThU} claims an uniqueness result but does not suggest a method of extrapolation from the set from $\I$.
Some linear predictors allowing  the required extrapolation can be found in \xxxonly{\cite{D17} and} \cite{D08}.
\end{remark}

The following corollary represents a modification for  processes of the classical sampling theorem
(Nyquist-Shannon-Kotelnikov Theorem). This Lemma states that
a band-limited function $x\in L_2(\R)$ is uniquely defined by the  sequence $\{x(t_k)\}_{k\in \ZZ}$, where
 given that
$X\ew=0$ for $\o\notin (-\O,\O)$ and $X=\F x$, $t_k=\tau k$; this theorem allows $\tau\in (0, \le \pi/\O]$.
There is  a version of this theorem  for oversampling sequences  with $\tau \in (0,\pi/\O)$: for any $s\in\ZZ$,  this $x$  is uniquely defined by the  sequence $\{x(t_k)\}_{k\in \ZZ, k\le s}$ \citet{F91,V87}.
Corollary \ref{corrS} below extends this version on the sampling Lemma in the case of   $\TT$-branched processes.

\begin{corollary}\label{corrS} Let the assumptions of Lemma \ref{ThU} be satisfied,  let $\O>0$ and $\tau \in (0,\pi/\O)$
be given, and let
$G_1=\R\setminus[-\O,\O]$ (i.e., the process $x_1$ is band-limited).
Then, for any $s\in\ZZ$,  the   $\TT$-branched process
$\{x_d\}_{d=1}^{m}$
is uniquely  defined (up to equivalency) by the sampling sequence  $\{x_1(t_k)\}_{k\in\ZZ,\ k\le s}$, where $t_k=\tau k$.
\end{corollary}

\begin{remark} For  $k>1$, the processes $x_k$ in  Corollary \ref{corrS}  are not necessarily band-limited.
Moreover, the sampling rate $\tau$   here does not depend
on the size of spectrum gaps $G_k$ of branches $x_k$ for $k\ge 2$.
This sampling rate depends only on the size of spectrum support for the single component $x_1$.
\end{remark}

To proceed further,  we introduce some additional conditions for  sets  $\TT=\{\G,I,H\}\in\TTT$
 restricting choices of $H$.

 For $j\in\{1,...,m\}$, let $A(j)\defi \{k:\ j\eqq k\}$. For a set $M\subset\{1,...,m\}$, let
$A(M)\defi \cup_{j\in M}A(j)$.

Staring from now, we assume that at least one of the following two conditions is satisfied.
\begin{condition} \label{condA0}
For any $(d,k)\in\G$, the operator  $h_{d,k}$ is the identity, i.e. $h_{dk}(x)=x$.
\end{condition}
\begin{condition} \label{condA}
There exists $n\in\{2,...,m\}$,  mutually disjoint subsets $M_p\subset\{2,...,m\}$,
$p=1,...,n$, and an open  set $D\subset \R$ of a positive measure
 such that the following holds.
\begin{enumerate}
\item $1\eqq d$ for all $d\in M_p$ for all $p=1,...,n$;
\item The sets $A(M_p)$ are mutually disjoint for $p=1,...,n$.
\item  If  $(d,k)\in\G$  and $d\in A(M_p)$ for some $p$, then $k\in A(M_p)$.
\item If $(d,k)\in\G$, then either $k\in \cup_{p=1}^n M_p$ or $k\in \cup_{p=1}^nA(M_p)$.
\item $h_{d,k}(x)=x$ for all $(d,k)\in\G$  such that  either $d\neq 1$ or
$d=1$ and $k \notin M_p$.
\item  For any  $p=1,...,n$, there exists an operator ${\rm h}_p:L_2(\R)\to L_2(\R)$
such that
\subitem(a)   $h_{1,k}(x)={\rm h}_p(x)$ for all  $k \in M_p$;
\subitem(b) $\F ({\rm h}_p x)=\HH_p(i\o)X(i\o)$ for $(d,k)\in\G_p$,
where $\HH_p\in L_\infty(i\R)$ is such that $\esssup_{\o\in D} |\HH_p(i\o)^{-1}|<+\infty$.
\end{enumerate}
\end{condition}

In particular, Condition \ref{condA} (vi)(b) holds if  $({\rm h}_{p}x)(t)=a x(b t+c)$ for some $a,b,c\in\R$, $b\neq 0$,
or if $({\rm h}_{p}x)(t)=\int_\R h(t-s)x(s)ds$, for some $h\in L_2(\R)$.

\begin{example}
The set $\TT$ in Example \ref{exA} satisfies Condition \ref{condA}
with
\baaa
&& M_1=\{2\},\quad  M_2=\{3\},\quad M_3=\{6,7\},\quad \breakk
A(M_1)=\emptyset,\quad  A(M_2)=\{ 4,5\},\quad
 A(M_3)=\emptyset.\eaaa
\end{example}
\begin{example}\label{exA1} {\rm  Consider processes defined in the time domain structured with
a  closed loop that have just two branches  $x_1$ and $x_2$. These branches are
connected  via restrictions that  $x_1(t)=x_2(t)$ for $t<0$,  and  $x_1(t)=x_2(t-1)$ for $t>1$. These processes
can be represented as  $\TT$-branched processes  $\{x_d(t)\}_{d=1}^2$ with
$\TT=(\{(1,2),\}, I,h)$, where $I_{1,2}=(-\infty,0)\cup(1,+\infty)$, and with operator
$h_{1,2}:L_2(\R)\to \R$ defined so that
$(h_{1,2}x)(t)=x(t)$ for $t<0$, and $(h_{1,2}x)(t)=x(1-t)$ for $t\ge 0$.
This $\TT$  does not satisfy Conditions \ref{condA0} and \ref{condA} since the operator
$h$ does not satisfy Condition \ref{condA} (vi)(b).  }
\end{example}

\begin{example}\label{exA2} {\rm
Technically,  Example \ref{exA1} can be modified so that the
same  restrictions for $x_1$ and $x_2$  hold  but  the operator
$h$ satisfy   Condition \ref{condA} (vi)(b). This can be achieved by adding a dummy branch.
More precisely, consider a $\TT$-branched process $\{x_d(t)\}_{d=1}^3$ with
$\TT=(\{(1,2),(3,1),(2,3)\}, I,h)$, where
\baaa
&&I_{1,2}=I_{3,1}=(-\infty,0), \quad I_{2,3}=(0,+\infty),\quad\breakk
h_{1,2}x=h_{3,1}x\equiv x,\quad  (h_{3,2}x)(t)=x(1-t).
\eaaa
Here  $x_3$ is a dummy branch supplementing the process from Example \ref{exA1}.
This corresponds to
 restrictions $x_1(t)=x_2(t)$ for $t<0$,  $x_1(t)=x_3(t)$ for $t<0$,
 $x_2(t)=x_3(1-t)$ for $t>1$. With this modification,  Condition \ref{condA}(vi) for $h$ is satisfied.
 However,  Condition\ref{condA} on $(\G,I)$ is  not satisfied.
}
\end{example}

 \begin{lemma}\label{ThDense} Let  $\TT=(\G,I,h)\in\TTT$ be such  that either Condition \ref{condA0} or  Condition \ref{condA} holds.  Then the following holds.
\begin{enumerate}
\item
 For any  $\TT$-branched process
$\{x_d\}_{d=1}^{m}\in \L_{2,\TT}$, and for any $\e>0$, there exists
$G\in\GG$ and a
$\TT$-branched process
$\{\w x_d\}_{d=1}^{m}\in \L^G_{2,\TT}$  such that
\baa
\max_{d=1,...,m}\|x_d-\w x_d\|_{ L_2(\R)}\le \e.
\label{xdd}\eaa
\item
For any branching  $\TT$-branched process
$\{x_d\}_{d=1}^{m}\in\CC_{\TT}$
 and any $\e>0$, there exists $G\in\GG$ and  a  $\TT$-branched process
$\{\w x_d\}_{d=1}^{m}\in \L_{2,\TT}^G\cap \CC_{\TT}$  such that \baa
\max_{d=1,...,m}(\|x_d-\w x_d\|_{ L_2(\R)}+\|x_d-\w x_d\|_{ C(\R)})\le \e.
\label{xdd2}\eaa
\item Let $\oo G_1\in \I$ be given such that
$\mes(D\setminus \oo G_1)>0$, where $D$ is the set in Condition \ref{condA}.   Let $x_1$ be such that $X_1(i\o)|_{\o\in\oo G_1} =0$ for $X_1=\F x_1$.
In this case, $G=(G_1,...,G_m)$ in statements (i) and (ii)  above can be selected so that $G_1=\oo G_1$.
\end{enumerate}
\end{lemma}

It should be emphasized  that  Lemma \ref{ThDense} claims  existence of processes $\{x_d\}$ featuring preselected spectrum degeneracy for the branches and, at the same time, such that the branches
coinciding  on preselected intervals.  This cannot be achieved by a simple application of
low/high pass filters to separate branches; if one applies such filters to $x_d$ and $x_k$,
this will impact the values on $I_{k,d}$, and the identity $x_k|_{I_{d,k}}=h_{d,k}(x_d)|_{I_{d,k}}$ could  be
disrupted.

 Lemma \ref{ThDense} combined
with Lemma \ref{ThU} allows to approximate a $\TT$-branched process by a process that can be recovered
from a single branch.  However, condition (\ref{IG})  in Lemma \ref{ThU} and Conditions
\ref{condA0}-\ref{condA} restrict choices of $\TT$ and $G$ where this approximation is feasible.\rd{
since if $I_{d,k}\notin\I_\infty$, $G_d\notin \I_\infty$, and $G_k\notin \I_\infty$, then $x_d\equiv x_k$.?}
Nevertheless,
there are choices of topology $\TT$ satisfying these conditions.

 The following result
provides sufficient conditions that ensure that a $\TT$-branched process can be recovered from its branch.
\begin{theorem}\label{ThM} Assume that $\TT=\{\G,I,H\}\in\TTT$ is such that
 either Condition \ref{condA0}  or  Condition \ref{condA} holds, and that
  $I_{d,k}\in \I_\infty$ for all $\{d,k\}\in\G$.
Then the following holds.
\begin{enumerate}
\item
 For  any  $\TT$-branched process
$\{x_d\}_{d=1}^{m}\in \L_{2,\TT}$
there exists $G\in \GG$ and a $\TT$-branched process
$\{\w x_d\}_{d=1}^{m}\in \L_{2,\TT}^G$ satisfying the assumptions of Lemma \ref{ThU} and such that
(\ref{xdd}) holds.
\item
For  any  $\TT$-branched process  $\{x_d\}_{d=1}^{m}\in \CC_{\TT}$
there exists $G\in \GG$ and a $\TT$-branched process
$\{\w x_d\}_{d=1}^{m}\in \L_{\TT}^G\cap \CC_{\TT}$ satisfying the assumptions of Lemma \ref{ThU} and such that
 (\ref{xdd2}) hold.
\end{enumerate}
The processes $\{\w x_d\}_{d=1}^{m}\in \L_{2,\TT}^G$  are uniquely defined by their path
$\w x_1|_{\oo I}$, for any $\oo I\in\I_\infty$. Under the assumptions  of Lemma \ref{ThDense}(iii) with $\oo G=G_1\in\I_\infty$, the processes $\{\w x_d\}_{d=1}^{m}$  are uniquely defined by the path
$\w x_1|_{\oo I}$ for any $\oo I\in\I$.
\end{theorem}

It can be noted that, under the assumptions of Theorem \ref{ThM}, by Corollary \ref{lemmaU2},
the spectrum gaps for different branch processes $x_d$ should be disjoint; otherwise,
the branches coincide, and it would makes some branches redundant in the model. This reduces choices of topological structures for processes   that can be recovered from a singe branch, since
the processes with compact spectrum gap can be recovered uniquely  from semi-infinite intervals of observations only.    However, there is an important example
that satisfy these restrictions such as examples listed below.

\begin{example}\label{ex3} {\rm
Consider
$\TT=(\{(1,d)\}_{d=1}^m, I,h)$, where $I_{1,d}=(-\infty,0)\cup (1,+\infty)$, and
where $h$ is any operator satisfying  Condition \ref{condA}.
This $\TT$  satisfy the assumptions of  Theorem \ref{ThM}.}
\end{example}
\begin{example}\label{ex4} {\rm
Consider
$\TT=(\{(1,d)\}_{d=1}^m, I,h)$, where $I_{1,d}=(-\infty,0)\cup (1,+\infty)$,
i.e.,  with   $x_1(t)=x_d(t)$ for $t\notin[0,1]$ for all $d$ for $\TT$-branched processes. This process satisfy the assumptions of  Theorem \ref{ThM}.}
\end{example}
\begin{example}\label{ex5} {\rm $\TT$
from Example \ref{exA}  satisfy the assumptions of   Theorem \ref{ThM} given that $\mes(G_5)=+\infty$.
On the other hand,  the sets $\TT$
from Examples \ref{exA1}-\ref{exA2} do not satisfy any of Conditions \ref{condA0} and \ref{condA}, and, respectively,
they do not satisfy the assumptions of Theorem \ref{ThM}.
}
\end{example}
\section{Applications: sampling theorem for branching processes}\label{SecMM}
\begin{theorem}\label{corrS2}  Let the assumptions of Theorem \ref{ThM}(ii) hold,  let $\O>0$ and  $\tau \in (0,\pi/\O)$
be given. \rd{, and let $G_1=\R\setminus[-\O,\O]$.}
Consider a $\TT$-branched process
$\{\ww x_d\}_{d=1}^{m}\in \L_{2,\TT}^G$ such that $X_1(i\o)=0$ if $|\o|>\O$ for $X=\F x$ (i.e., the process $x_1$ is band-limited).
Then, for any $\e>0$,   there exists   $G\in\GG$ and
$\TT$-branched process $\{\ww x_d\}_{d=1}^{m}\in \L_{2,\TT}^G\cap \CC_\TT$
such that
$G_1=\R\setminus[-\O,\O]$, that (\ref{xdd}) holds, and that
the following holds:
\begin{enumerate}
\item
 the  $\TT$-branched process
$\{\ww x_d\}_{d=1}^{m}\in \L_{2,\TT}^G$
is uniquely  defined  (up to equivalency) by the sampling sequence  $\{x_d(t_k)\}_{k\in\ZZ,d=1,..,m}$, where $t_k=\tau k$.
\item
Moreover,  for any
 $s\in\ZZ$,  the $\TT$-branched process
$\{\ww x_d\}_{d=1}^{m}\in \L_{2,\TT}^G$
is uniquely  defined  (up to equivalency) by the sampling sequence  $\{x_1(t_k)\}_{k\in\ZZ,\ k\le s}$, where $t_k=\tau k$.
\end{enumerate}
\end{theorem}

The conditions of Theorem \ref{corrS2} restrict choices of branched processes. However, they still hold for many models
describing  branching scenarios.\begin{example}\label{ex2} {\rm Let $x_1(t)$ be   a coordinate of a fighter jet tracked by a locator for time $t<0$, and let
this jet ejects $m-1$ false targets at time $t=0$; these false targets move  according to different evolution laws.
This can be modelled by a $\TT$-branched process $\{x_d\}_{d=1}^m$  with a
$\TT=(\{(1,d)\}_{d=2}^m,I,h)$, where $I_{1,d}=(-\infty,0)$, and where $h_{1,d}(x)\equiv x$,
i.e. with   $x_1(t)=x_d(t)$ for $t<0$ for all $d$.  It is not obvious how to apply  the approach of the classical sampling theorem
in this situation.  On the other hand, this case is covered by Theorem \ref{corrS2}. Therefore,  for any $\e>0$, there exist  a $\TT$-branched process $\{\w x_{d,\e}\}_{k=1}^m$,  such that the following holds:
\begin{enumerate}
\item
$\sup_{t\in\R}|\w x_{1,\e}(t)-x_1(t)|\le\e$, \quad $\sup_{t> 0}|\w x_{d,\e}(t)-x_d(t)|\le\e$;
\item
For any $\tau\in (0,\pi/\O)$ and  $s<0$,  an  equidistant  sequence $\{\w x_{\e,1}(\tau k)\}_{k\in\ZZ,\ k<s}$ defines $\{\w x_{d,\e}(\cdot)\}_{d=1}^m$ uniquely. \end{enumerate}}
\end{example}
\section{Proofs}\label{SecP}

{\em Proof  of Proposition \ref{lemmaU}}. The statements of this proposition  are known;  for completeness, we provide the proof.

Cleary,   $\mes(\oo I\cup \oo G)=\infty$  if and only if either $\oo I\in\I_\infty$ or $\oo G\in\I_\infty$.
Let us consider the case where $\oo I\in\I_\infty$.   Without loss of generality, we assume that
 $(-\infty,0)\subset\oo I$. Let
$\C^+\defi\{z\in\C:\ \Re z>  0\}$, and let  $H^2$ be the Hardy space of holomorphic on $\C^+$ functions
$h(p)$ with finite norm
$\|h\|_{H^2}=\sup_{s>0}\|h(s+i\o)\|_{L_2(\R)}$; see, e.g. \cite{Du}, Chapter 11.
 It suffices to
prove that if $x\in L_2(\R)$ is such that $X\ew =0$ for $\o\in\oo G$, $X=\F x$,  and  $x(t)=0$ for
$t\le 0$,  then  $x(t)=0$ for $t>0$. These properties imply that $X\in H^2$, and, at the same time,
\baaa \int_{-\infty}^{+\infty}(1+\o^2)^{-1}|\log|X\ew||dx=+\infty.
\eaaa
 Hence, by the property of the Hardy space,
  $X\equiv 0$; see, e.g.  Lemma 11.6 in \cite{Du}, p. 193.
This proves the statement of Proposition \ref{lemmaU} for the case where
$\oo I\in\I_\infty$.  Because of the duality between processes in time domain and their Fourier transforms,
this also implies the proof for the case where $\oo G\in\I_\infty$. This completes the proof of Proposition  \ref{lemmaU}. $\Box$

{\em Proof  of Corollary \ref{lemmaU2}} follows immediately from the definitions
and from Proposition \ref{lemmaU}. $\Box$

{\em Proof  of Lemma \ref{ThU}}. Let $\oo I\in \I$ be such that $\oo I=\R$ for statement (i),
 $I\in I_\infty$ for statement (ii),
 $I\in \I$ for statement (iii).

 By Proposition \ref{lemmaU},  $x_1$ is uniquely
defined by $x_1|_{\oo I}$. Further, let $N=d_N\in\{2,3,...,m\}$ be given.  By Lemma \ref{lemmaU},  $x_{d_1}$ is uniquely
defined by  $h_{d_1,d_0}(x_{1})|_{I_{d_0,d_1}}$, i.e., by  $x_1|_{\oo I}$.  Similarly, $x_{d_2}$ is uniquely
defined by $x_{d_2}|_{I_{d_2,d_1}}  =h_{d_1,d_2}(x_{d_1})|_{I_{d_2,d_1}}$,  i.e., by  $x_1|_{\oo I}$ again.
Repeating  this for all $d_k$, $k=1,...,j$, we obtain that $x_{d_j}$ is uniquely
defined by $x_{1}|_{\oo I}$.  Hence  $x_{d}$ is uniquely
defined by $x_{1}|_{\oo I}$.
This completes the proof of Lemma \ref{ThU}. $\Box$

{\em Proof  of Corollary \ref{corrS}}. It follows from  the results \citet{F91,V87}
that $x_1$ is  uniquely  defined by $\{x_1(t_k)\}_{k\le s}$. Then the statement of Corollary \ref{corrS}
follows from Lemma \ref{ThU}. $\Box$

{\em Proof of Lemma \ref{ThDense}}.  Let us suggest a procedure for the construction of $\w x$; this will be sufficient to prove the theorem. This procedure  is given below.
\par
\vspace{2mm}
Let us assume first that  Condition \ref{condA} holds.
\par
For $(d,k)\in\G$, let $H_{kd}(i\o)=1$ if either  $d\neq 1$ or $k\notin \cup_{p=1}^n M_p$, and where
 $H_{1 k}(i\o)={\rm H}_p(i\o)$ if $k\in M_p$.

 Let \baaa y_{k,d}\defi x_{k}-h_{dk}(x_{d}),\quad Y_{k,d}\defi \F y_{k,d}= X_k-h_{kd}X_d,
 \eaaa
 where \bl{ $ X_{k}\defi \F  x_{k}$.}

Consider  a set $\{\o_{k}\}_{k=1,...,m}\subset \R$ such that $\o_k$ are located in the interior $D\setminus \oo G_1$ for $k\ge 2$.
 Let $G_k=J_{k}(\d)$.
Here $J_k(\d)\defi (\o_{k}-\d,\o_{k}+\d)$ for $k>1$,
 \bl{$J_{1}(\d)\defi (\o_{1}-\d,\o_{1}+\d)$ } if $G_1$ has to be selected, and
 $J_{1}(\d)\defi \oo G_1$ \rd{\}} if $G_1=\oo G_1$ is fixed (i.e., in  the case of Lemma \ref{ThDense}(iii)).

 We assume below that $\d>0$ is small enough  such that these intervals are disjoint and that
 $J_k(\d)\subset D$ for $k\ge 2$;
 this choice of $\d$ is possible since   $\o_k\neq \o_j$ if $j\neq k$.

Let $M^c\defi \{k=2,...,m\}\setminus (\cup_{p=1}^n M_p)$ and
$B\defi  M^c\cup (\cup_{p=1}^n A(M_p))$.   Set
\baaa
\w X_1\ew\defi X_1\ew\Ind_{\{\o\notin \cup_{d=1}^{m} J_{d}(\d)\}} -
\sum_{d\in B} Y_{d,1}\ew\Ind_{\{\o\in J_{d}(\d)\}}
\\-\sum_{p=1}^n \HH_p(i\o)^{-1}\sum_{d\in M_p} Y_{d,1}\ew\Ind_{\{\o\in J_{d}(\d)\}}
 \eaaa
and
\baaa
\w X_d\defi H_{1,d}\w X_1+ Y_{d,1}, \quad d=2,...,m.
\eaaa
\par
For $k\in M_p\cup A(M_p)$ and $d\in A(M_p)$, we have that
\baaa
\w X_k-H_{d,k}\w X_d=\w X_k-\w X_d=  \w X_1+Y_{k,1}-\w X_1-Y_{d,1}
\brea=Y_{k,1}- Y_{d,1}=X_{k}-X_1-X_{d}+X_1=Y_{k,d},
\eaaa
i.e.
\baaa
\w X_k=\w X_d+Y_{k,d}.
\eaaa
Let $\w x_d=\F^{-1}\w X_d$, $d=1,...,m$.

Under the assumptions of statement (i) of the theorem, we have that  $x_{k}|_{I_{d,k}}=H_{d,k}(x_{d})|_{I_{d,k}}$ up to equivalency. It follows that $y_{k,d}|_{I_{d,k}}=0$ up to equivalency, i.e.
$\w x_k|_{I_{d,k}}=h_{d,k}(\w x_d)|_{I_{d,k}}$ up to equivalency.
Since this holds for all   $(d,k)\in\G$, it follows that $\{\w x_d\}_{d=1}^m$ is a   $\TT$-branched process with the same structure set $\TT$ as the underlying   $\TT$-branched process $\{ x_d\}_{d=1}^m$.

Let us show  that the   $\TT$-branched process
$\{\w x_d\}_{d=1}^{m}$ features the required  spectrum degeneracy.

Since the intervals $J_d(\d)$ are mutually disjoint,
 it follows immediately from the definition for $\w X_1$ that   $\w X_1\ew =0$ for $\o\in J_1(\d)$.

Further, by the definition for $\w X_d$ for $d>1$, we have that
\baaa
&&\w X_d\ew=H_{1,d}\Bigl(X_1\ew\Ind_{\{\o\notin \cup_{d=1}^{m} J_{d}(\d)\}}\breakk  -
\sum_{d\in B} Y_{d,1}\ew\Ind_{\{\o\in J_{d}(\d)\}}
\\&&-\sum_{p=1}^n \HH_p(i\o)^{-1}\sum_{d\in M_p} Y_{d,1}\ew\Ind_{\{\o\in J_{d}(\d)\}}\Bigr)
+Y_{d,1}\ew.
\eaaa
Since the intervals $G_d=J_d(\d)$ are mutually disjoint, we have that
\baaa
&& \w X_d \ew \Ind_{\{\o\in J_d(\d)\}}\breakk= 0-
Y_{d,1}\ew\Ind_{\{\o\in J_d(\d)\}}+Y_{d,1} \Ind_{\{\o\in J_d(\d)\}}=0.
 \eaaa
We obtain that separately for $d\in B$ and $d\notin B$, using properties of $\HH_p$, $H_{1,d}$
implied  from their definitions.

It follows that $\w X_d\ew =0$ for $\o\in J_d(\d)$ for $d>1$ as well.  It follows that the    $\TT$-branched process
$\{\w x_d\}_{d=1}^{m}$ belongs to $\L_{2,\TT}^G$ with $G_d=J_\d(d)$, i.e. features the required spectrum degeneracy.

Furthermore, for all $d$,  we have that, under the assumptions of statement (i),
$\|X_d(i\cdot)-\w X_d(i\cdot)\|_{L_2(\R)}\to 0$ as $\d\to 0$.
In addition,    we have that, under the assumptions of statement (ii),
 \baaa
 \|X_d(i\cdot)-\w X_d(i\cdot)\|_{L_2(\R)}+\|X_d(i\cdot)-\w X_d(i\cdot)\|_{L_1(\R)} \to 0\eaaa
 as $\d\to 0$. Under the assumptions of statement (i), it follows that $\|\w x_d-x_d\|_{ L_2(\R)}\to 0$. Under the assumptions of statement (ii), it follows that  $\|\w x_d-x_d\|_{ L_2(\R)}+\|\w x_d-x_d\|_{ L_2(\R)}\to 0$  as $\d\to 0$.
 \par
 \vspace{2mm}
For the proof under Condition \ref{condA0}, we select
 \baaa
\w X_1\ew\defi X_1\ew\Ind_{\{\o\notin \cup_{d=1}^{m} J_{d}(\d)\}} -
\sum_{d=2}^m Y_{d,1}\ew\Ind_{\{\o\in J_{d}(\d)\}}.
\eaaa
Then the proof is similar to the  proof for the case where Condition \ref{condA} holds.
This completes the proof of Lemma \ref{ThDense}. $\Box$

It can be noted that the construction in the proof of Lemma \ref{ThDense}  follows the approach suggested in \cite{D19} for discrete time processes.
Let us illustrate the construction using a toy example.
\begin{example} {\rm Let $m=2$, and $\TT=(\G,I,h)$
be such that
$\G=\{(1,2)\}$, $I_{1,2}(x)=(-\infty,0)\cup (1,+\infty)$,  $h_{1,2}(x)=x$.
This choice imposes restrictions $x_1(t)=x_2(t)$ for $t\notin [0,1]$.

Further,  in the notations of the  proof of Lemma \ref{ThDense}(iii),  let $\oo G_1=\{\o\in\R:\ |\o|>1\}$, $\o_2=0$,
$x_1(t)\equiv 0$, and $x_2(t)=\Ind_{[0,1]}$.  In this case,  we have that
\baaa
X_1(i\o)= 0,\quad  X_2(i\o)=\frac{1-e^{-i\o}}{i\o},\quad  Y_{2,1}(i\o)=X_2(i\o).\eaaa
Let us select $J_1(\d)=\oo G_1$ and $J_2(\d)=\{\o:\ |\o|\le \d\}$, $\d\in(0,1)$. The corresponding processes $\w X_d$
are
\baaa
&&
\w X_1(i\o)=0- Y_2(i\o)\Ind_{\{|\o|\le \d\}}=- X_2(i\o)\Ind_{\{|\o|\le \d\}}  ,\quad\breakk
\w X_2(i\o)=\w X_1+Y_{2,1}=X_2(i\o)\Ind_{\{|\o|> \d\}}.
\eaaa
This gives
\baaa
&&\w x_1(t)=\frac{1}{2\pi}\int_{-\d}^\d e^{i\o} \frac{1-e^{-i\o}}{i\o}d\o,\breakk
\w x_2(t)=x_2(t)+\ww x_1(t).
\eaaa
Clearly,  $\w x_1(t)=\w x_2(t)$ for $t\notin [0,1]$. Hence $\{\w x_d\}_{d=1,2}\in \L_{2,\TT}^G$ is a $\TT$-branched process with
$G=(\oo G_1,G_2)$ and $G_2=\{\o\in\R:\ |\o|\le \d\}$.
For sufficiently small $\d$, the processes $\w x_d$ can be arbitrarily close to $x_d$.
The process $\w x_2$ has a spectrum gap $J_2(\d)$ and can be recovered \cite{D08} from its path $\w x_2|_{t<0}=\w x_1|_{t<0}$; this recovery is uniquely defined  in the class
of processes featuring this spectrum gap.
The process $\w x_1$ is band-limited  and can be recovered from its semi-infinite sample as described in Corollary \ref{corrS}; this recovery is uniquely defined  in the class
of band-limited processes with the same spectrum band.
}\end{example}

{\em Proof  of Theorems \ref{ThM} and \ref{corrS2}} follows immediately from  Lemmas \ref{ThU}--\ref{ThDense}. $\Box$

\section{Conclusions and future research}\label{SecC}
The present paper is focused on the frequency analysis  for
processes
with time domain represented as oriented branched 1-manifolds that can be considered as an oriented graph with continuous
connected branches.
The paper suggests an approach that allows to take into account
the topology of the branching line via modelling it  as a system of standard processes  defined on the real axis and
coinciding on preselected intervals with well-defined Fourier transforms (Definition \ref{def1}).
This approach allows  a relatively simple and convenient
representation of processes defined on time domains represented as a 1-manifold, including manifolds represented by restrictions    such as $x_k(t)=x_d(t+\tau)$ or
$x_k(t)= x_d(\tau-t)+c$, or $x_k(t)=\int_\R h(t-s)x_d(s)ds$, for $t\in I$, with  arbitrarily chosen  preselected  $I_{d,k}\subset \R$, $c,\tau\in\R$, and  $h\in L_2(\R)$.

It could be interesting to extend the results on processes with  time domain represented as compact  oriented branched 1-manifolds.  Possibly, it can be achieved via extension of the domain of these processes.
For example,  one could  extend
edges of compact branching line beyond  their vertices and  transform finite edges into semi-infinite ones.
Alternatively,  one could supplement the branching lines by new dummy semi-infinite edges originated from the vertices of order one. We leave them for the future research.

\bl{In addition, Lemma \ref{ThU} and Theorem \ref{ThM}   can be extended on the case of processes $\ww x_d$ with the spectrum vanishing at single points with sufficiently high vanishing rate
such as described in \cite{D21}.}


\end{document}